\documentclass[12pt]{article}
\usepackage{epsfig,amsmath}
\usepackage[usenames]{color}
\usepackage{achemso}
\setkeys{acs}{usetitle = true}
\usepackage[sort&compress,numbers,super]{natbib}
\usepackage{comment}
\usepackage{colortbl}
\usepackage{tabu}
\usepackage[usenames,dvipsnames]{xcolor}
\usepackage[normalem]{ulem}
\usepackage{float}
\usepackage{caption}
\usepackage{subfig}
\usepackage{graphicx}
\usepackage{hyperref}
\usepackage[utf8x]{inputenc}
\usepackage{enumitem}

\begin{document}

\begin{center}

{\large\bf{A Review on Parallel Virtual Screening Softwares for High Performance Computers}}

\vspace{0.3cm}
{\bf Natarajan~Arul~Murugan$^{\dagger}$*, Artur~Podobas$^{\dagger}$, Davide~Gadioli$^{\ddagger}$, Emanuele~Vitali$^{\ddagger}$, Gianluca~Palermo$^{\ddagger}$ and Stefano~Markidis$^{\dagger}$*}\\

$\dagger$ \quad Department of Computer Science, School of Electrical Engineering and Computer Science, KTH Royal Institute of Technology, Stockholm, SE-10044, Sweden;  

$^\ddagger$ \quad Dipartimento di Elettronica, Infomazione e Bioingegneria, Politecnico di Milano, Milano 20133, Italy.

\end{center}

\begin{abstract}
Drug discovery is the most expensive, time demanding and challenging project in biopharmaceutical companies which aims at the identification and optimization of lead compounds from large-sized chemical libraries. The lead compounds should have high affinity binding and specificity for a target associated with a disease and in addition they should have favorable pharmacodynamic and pharmacokinetic properties (grouped as ADMET properties).
Overall, drug discovery is a multivariable optimization and can be carried out in supercomputers using a reliable scoring function which is a measure of binding affinity or inhibition potential of the drug-like compound. The major problem is that the number of compounds in the chemical spaces is huge making the computational drug discovery very demanding. However, it is cheaper and less time consuming when compared to experimental high throughput screening. As the problem is to find the most stable (global) minima for numerous protein-ligand complexes (at the order of 10$^6$ to 10$^{12}$), the parallel implementation of in-silico virtual screening can be exploited to make the drug discovery in affordable time.
In this review, we discuss such implementations of parallelization algorithms in virtual screening programs.  The nature of different scoring functions and search algorithms are discussed, together with a performance analysis of several docking softwares ported on  high-performance computing architectures.
\end{abstract}

\vspace{2cm}

\section{Introduction}
\label{intro_sec}
Drug discovery is one of the highly challenging, time consuming and the most expensive projects in the healthcare sector. The usual time involved in 
 bringing a drug from basic research to market is 12-16 years and the cost associated is about 2.5 billion dollars~\citep{para_vs1,para_vs2,para_vs3,para_vs4}. To meet one of the EU sustainable development goals~\cite{sdg17} aimed at the good health and well-being for everyone, drugs should be made available to the common people in an affordable price and the current protocols in drug development need to be redesigned to make the discovery process economically sustainable. One of the most promising technique to accelerate the drug discovery process, and to make it more cost-effective, is to perform in-silico virtual screening, and to exploit the computational power of large High-Performance Computing (HPC) systems.

One of the major contributing factors to the cost and time associated with the discovery
is that it has been reported~\citep{petrova2014innovation,kiriiri2020exploring} that only one in 10,000 compounds subjected to research and development (R\&D) comes out to be successful. The drug discovery involves various steps such as target discovery, lead identification, lead optimization, ADMET (Absorption, Distribution, Metabolism, Excretion, Toxicity) properties optimization, and clinical trials~\cite{para_vs5}. Once a valid target is known for a disease,  compounds from different chemical libraries are subjected to high-throughput screening against this target. If the number of compounds used for screening can be narrowed down to a few hundreds, the cost and time associated with a drug discovery process can be drastically reduced. Using computational approaches, many of the steps involved in the drug discovery projects can be made cost effective and less time consuming. For example, in the case of protein tyrosine phosphatase-1B~\cite{para_vs6}, the experimental high-throughput screening of a chemical library with 400,000 compounds yielded a success rate of 0.021\% in identifying the ligands that can inhibit the enzyme with IC$_{50}$ values less than 100 $\mu$M. However, with the use of a preliminary screening phase using a computational approach, the success rate came out to be 34.8\% starting from a chemical library of 235,000 compounds.

To summarize, the experimental high throughput screening is not suitable to deal with modern chemical spaces since they are composed up to billions of molecules. To solve this problem, it is common to use computational approaches on HPC systems. 
In this review, we highlight various currently available implementations of virtual screening softwares suitable for high performance computers. Below, we provide general introduction to virtual screening (VS) problem and discuss about the possibilities for the parallelization so that it can be effectively implemented for computing facilities offered by HPCs. 
 
 The paper is organized as follows. Section~\ref{vs_sec} introduces the computational VS with details on scoring functions and search algorithms. Section~\ref{algorithms} presents details on the major breakthroughs obtained in VS and Section~\ref{miles_sec} presents the main parallelization techniques used in VS and why they target HP systems. In Section~\ref{hpc_sec}, we provide an overview about the implementations of different VS softwares. Finally, we discuss about the opportunities offered by reconfigurable architectures such as FPGAs.   
 
\section{The In-silico Virtual Screening Problem}
\label{vs_sec}
\begin{center}

\begin{figure}[h]
	\begin{center}
    \includegraphics[width=0.75\textwidth]{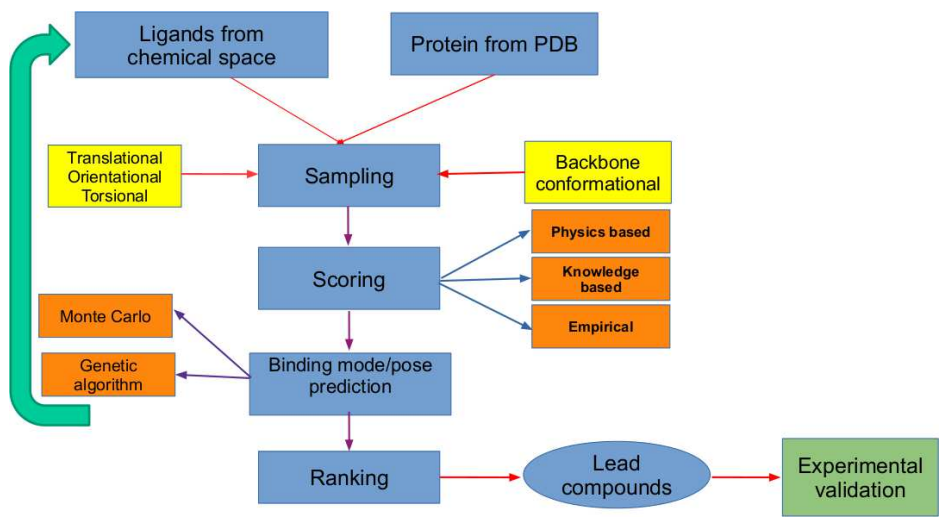}
    \caption{Workflow for computer aided drug discovery where the lead compounds are identified using VS. It shows that the various configurations of the ligand within binding site are generated using sampling which are scored using a scoring function to identify the most stable binding mode/pose. The docking energies of the most stable configurations of all the ligands  are used for ranking them to identify a list of lead compounds which are taken for further experimental validation. As the sampling and scoring need to be done for all the ligands in the chemical library, these steps are shown within a loop.} 
    \label{figure_cadd}
	\end{center}
\end{figure}
\end{center}

In general, the computational approaches for molecular docking have two main components: Sampling and Scoring. Sampling refers to generation of various conformations and orientations for the ligand within a target binding site (defined usually by a grid box). Scoring refers to evaluating the binding/docking energies for various configurations of the ligand within the binding site. The most stable configuration of the ligand is referred to as binding pose. The VS protocol where molecular dockings are carried out for all the ligands from a chemical library includes a third component referred to as ranking where different ligands are ranked with respect to their binding potential.  Overall, the VS identifies the ligand with the topmost binding affinity (which is based on the docking energies of different ligands) for a given biomolecular target. In addition the most stable binding mode/pose for each of the ligands within the binding site is found (which is based on the relative docking energies of different configurations of the same ligand). Figure \ref{figure_cadd} shows the general workflow of computer aided drug discovery where VS approach is used to identify lead compounds. It shows the steps involved in the binding pose identification of ligands within the binding site and subsequently the ranking of different ligands is carried out to identify the lead compounds. Most of the VS schemes do not include the flexibility for the target protein and only the sampling over translational, rotational, and torsional degrees of freedom of ligand is accounted for.

\subsection{Scoring Functions}
\label{sf_sec}
The reliability and accuracy of the scoring functions used for screening compounds are the most important parameters that dictate the success rate of the computational screening approaches. The scoring functions are mostly defined to be proportional to the binding affinity of the ligand towards a target. The scoring functions are often classified as \textit{physics-based}, \textit{knowledge-based} and \textit{empirical}. 

\begin{enumerate}[label=(\roman*)]

\item \textit{Physics-based} scoring functions are based on the binding free energies which are the sum of various interactions between protein-ligand subsystems such as van der Waals, electrostatic, hydrogen bonding, solvation energy and entropic contributions.

\item The \textit{knowledge-based} scoring functions are based on the available protein-ligand complex structural data from which the distributions of different atom-atom pairwise contacts are estimated. The frequency of appearance of different pairwise contacts are used to compute potential mean force which is used for ranking protein-ligand complexes. 

\item Finally, the \textit{empirical scoring} functions as the name implies are based on empirical fitting of binding affinity data to potential functions whose weights are computed using a reference test system. Modern scoring functions mainly fall in this class including the machine learning based approaches built based on the available information on the protein-ligand 3D structures and inhibition/dissociation constants~\cite{para_vs7}. 

\end{enumerate}

As we discussed above, there are different scoring functions developed and this section mainly focuses on implementations available in open source softwares such as Dock~\cite{dock1}, Autodock4.0~\cite{para_vs8}, Autodock Vina~\cite{para_vs9} and Gnina~\cite{para_vs10}.  
The docking energy defined to rank protein-ligand complexes ($sf$) in Autodock4.0 is classified as physics based and is defined as the sum of van der Waals, electrostatic, hydrogen bonding and desolvation energy, as shown in Equation~\ref{scoringfunc}. In addition, the entropic contribution which is proportional to number of rotatable bonds is also added to the docking energy. In the equation, r$_{ij}$ refers to the distance between the two atoms, i and j centered on protein and ligand subsystems. Similarly, q$_i$ and q$_j$ refer to charges on these atoms. A$_{ij}$,  B$_{ij}$ are the coefficients of the potential energy functions describing van der Waals interaction. C$_{ij}$,  D$_{ij}$ are the coefficients of the potential energy functions describing hydrogen bonding interaction. The terms S$_i$ and V$_i$ refer to the solvation parameter and fragmental volume of atom, i respectively.

\begin{equation}
\label{scoringfunc}
\scriptsize
sf=W_{vdw}\sum_{i,j} (\frac{A_{ij}}{r_{ij}^{12}}- \frac{B_{ij}}{r_{ij}^{6}})  + 
W_{HB}\sum_{i,j} (\frac{C_{ij}}{r_{ij}^{12}}- \frac{D_{ij}}{r_{ij}^{10}})  +
W_{elec}\sum_{i,j} (\frac{q_iq_j}{\epsilon(rij)r_{ij}}) +
W_{ds}\sum_{i,j}(S_iV_j+S_jV_i)  (exp\frac{-r_{ij}^2}{2\sigma^2}) 
\end{equation}

Since the protein-ligand complexes are considered to be in an aqueous environment, the binding free energies need to account for this and solvation energy adds the binding free energy differences due to vacuum to aqueous like environments. In particular, the last term in the equation accounts for this solvation effect (refer to Equation ~\ref{scoringfunc}).

In general, the entropic contributions can be due to translational, rotational and torsional degrees of freedom. However, the docking energies implemented in the aforementioned molecular docking softwares only account for the contributions due to torsional degrees of freedom. The contributions due to other degrees of freedom are assumed to be negligible in ranking different ligands. It is worth noting that certain free energy calculations tools such as MMPBSA.py estimate the entropic contributions due to all degrees of freedom based on normal mode analysis~\citep{miller2012mmpbsa}. 
The entropic contributions due to torsional degrees of freedoms in molecular docking softwares are oversimplified and they are estimated from the number of rotatable bonds (each bond contributes with 0.2-0.5 kcal/mol).~\citep{arul_wiley} In the case of Autodock Vina, the scoring function can be majorly classified as empirical in nature and it is a sum of various distance dependent pairwise interactions~\cite{para_vs9}.  It includes terms for describing steric, hydrophobic and hydrogen bond interactions. The values for different parameters and weights for different terms of potential functions were obtained from non-linear regression to PDBbind 2007 dataset. In other words, the empirical scoring functions can have the same mathematical expression as in equation 1 but the weights/coefficients for different types of interactions are obtained from fitting to experimental binding potentials. The knowledge based scoring functions~\citep{dittrich2018converging} (sf$\mathrm{_{kb}}$) have the following form:

\begin{equation}
\label{kb_sf}
sf_{kb}=    \sum_{i}^L\sum_{j}^R -k_BTln[g(r_{ij})]
\end{equation}

\noindent
Here the summation runs over the ligand and receptor atoms, and g(r$\mathrm{_{ij}}$) refers to the relative probability distribution of distances of a specific types of protein-ligand atom pairs in the docked complex structure when compared to reference experimental complex structure.

Recently deep-learning networks are proposed to provide scoring functions. For instance, Gnina uses convolutional neural network (CNN) based scoring function to rank the protein-ligand complexes~\cite{para_vs10}. The neural networks were trained 
using three-dimensional protein-ligand complex structures from the PDBbind database. In particular, the dataset contained two sets of poses for the ligands within the binding site. The group of positive poses which had root mean square deviation (RMSD) value below 2 Å  when compared to the crystallographic poses while the rest were considered as a group of negative poses. Here, RMSD is the root mean square deviation in atomic positions in the predicted pose when compared to reference pose as in the crystal structure. The positive and negative poses were generated using the experimental protein-ligand three dimensional structures by adopting a random conformation generation algorithm. The CNN model has been trained using the 4D grid (which was constructed using the protein-ligand coordinates within the grid box and atom types) to classify the poses. 

\subsection{Search Algorithms}
\label{algorithms}
The search algorithms aim at finding the protein-ligand structure corresponding to the global minimum in a potential energy surface. However, this is very challenging problem and many search algorithms end up in a local minima.
Therefore, molecular docking software use several techniques such as deterministic search \cite{beccari2013ligen}, Genetic algorithm, Monte Carlo with simulated annealing\citep{mc_sa}, particle swarm optimization\citep{particle_swarm}, or Broyden-Fletcher-Goldfarb-Shanno (BFGS)~\cite{para_vs9}. Deterministic approaches apply techniques such as gradient descend, and they focus on a reproducible sequence of pose evaluations.
Monte Carlo algorithms generate numerous poses by using random values for translational, rotational and torsional degrees of freedom of the ligand. In Monte Carlo simulated annealing, the heating step allows the system to escape from the local minimum (during which it can sample high energy regions of the potential energy surface).
In genetic algorithms each pose is defined by a vector of genes that correspond to translational, rotational and torsional degrees of freedom.
By varying these values in the genes, new poses can be generated.
The fitness function aims at finding the minimum energy of the pose.

\subsection{Validation of molecular docking approaches}
As we discussed above molecular docking approaches employ different types of scoring functions and before implementation they were validated rigorously against available experimental data. In particular, two properties obtained from molecular docking can be considered in general for benchmarking: 

\begin{enumerate}[label=(\roman*)]
\item RMSD computed for the predicted binding pose against the crystallographic pose obtained experimentally.
 \item Binding free energies/docking energies which are proportional to experimental inhibition/dissociation constants. 
 \end{enumerate}

The RMSD in the above list is computed from the experimental and predicted protein-ligand complex structures and provides an estimate about how well the molecular docking software is capable of producing the most stable binding mode and binding pose of the ligand within the target biomolecule. A RMSD value of $<$2 Å is considered as a threshold value for the correct prediction of complex structure~\cite{para_vs21}.

A benchmark study using Autodock4.0 and Autodock Vina on the complex structures (190 in number) from PDBbind showed that latter could predict structures within the threshold value (i.e. $<$2Å) for about 78\% complexes while the former one achieved 42\%~\cite{para_vs9}. There are other studies reported in the literature which compared the performance of various molecular docking softwares such as AutoDock, DOCK, FlexX, GOLD and ICM with the ICM came out to be the superior performer with structures predicted for 93\% complexes within the acceptable accuracy~\cite{para_vs21}. 

The other set of quantities used for benchmarking the molecular docking approaches are the inhibition constants, dissociation constants, IC$_{50}$ and pIC$_{50}$ which are available from experimental binding assay studies. All these quantities refer to the binding potential or inhibiting potential of ligands to a specific biomolecular target. The dissociation constants and binding free energies are related to each other through the following Equation:

\begin{equation}
\label{delta_g}
    \Delta G = RTlnK_d
\end{equation}

Where $R$ is the gas constant (equals to 1.987 cal K$^{-1}$ mol$^{-1}$) and $T$ refers to temperature (set to 298.15 K). Thanks to this equation, the computed docking energies can be directly validated using the experimental binding assay results.

\subsection{Computational cost associated with virtual screening}
In a computational drug discovery project high affinity lead compounds against a target biomolecule (can be an enzyme, membrane protein, DNA, RNA, Quadruplex, membrane or fibril aggregates) are identified using a reliable scoring function. One can identify lead compounds for a target from various chemical spaces. 
The popular chemical spaces are ZINC, Cambridge, Chemspider, ChEMBL, Pubchem, Pubmed, DrugBank, TCM, IMPPAT and GDB13-17. Refer to Table \ref{topchemical} for a list of different chemical libraries with their properties~\cite{walters2018virtual}. The estimate for the size of chemical space for carbon based compounds with molecular mass $<$500 Daltons is 10$^{60}$ which clearly indicates that there are limitless possibilities for designing a therapeutic compound. 
As can be seen, even the use of the top most supercomputers with exascale computing speed (which can do $10^{18}$ floating point operations per second running the High-Performance Linpack benchmark) for screening these compounds will take universal life time. So, it is reasonable to use the certain filters to reduce the number of compounds before subjecting to screening. Recently some filtering procedure based on Bayesian optimization algorithm referred to as MolPAL was used to identify top 50\% compounds by developing a model with data of explicit screening of only less than 5\% compounds of the chemical space~\cite{para_vs23}.  

Otherwise, it is reasonable to use other chemical libraries having compounds that are easy to synthesize and having favourable pharmacokinetic (ADMET) properties. The chemical library with the largest number of compounds is  GDB17  that contains 166 billion organic molecules made of just 17 atoms of C, O,N, S and halogens~\cite{para_vs24}. Most of the drug discovery applications use the DrugBank database, Enamine database, ZINC15,~\cite{para_vs25} Cambridge and the number of compounds in these chemical libraries are listed in Table~1. As can be seen the compounds range from tens of thousands to billions and the computational cost associated with screening is enormous which requires use of the HPCs and accelerators. 

 To demonstrate the computational demand associated with virtual screening we describe an application below. In the case of the AmpC target, Lyu et al. docked 99 million molecules. For each compound 4000 orientations on average were generated.Further for each orientation about 280 conformations were generated~\citep{para_vs16}. So, for each ligand, 1.1 M docking energy calculations were carried out and given that the number of compounds considered were 99 M, a total of 10$^{13}$ of such calculations were carried out. This will be further increased in the flexible receptor docking where the sampling over side chain conformations of residues needs to be accounted for~\citep{para_vs16}. As can be seen, the computational demand is really huge with such virtual screening applications and so it is inevitable to develop parallel algorithms and to use HPCs to accomplish such screenings within affordable time. 

\noindent
\begin{table}[]

\scriptsize
    \centering
     \caption{Top chemical spaces available for VS }
\begin{tabular}{lll} \hline
{\bf Chemical library} & {\bf NO of compounds} & {\bf Features}\\ \hline
Virtual compounds &10$^{60}$ & Molecular mass $\leq$500 Daltons \\
GDB17\citep{walters2018virtual} &166 B & 17 heavy atoms of type C,O,N,S and halogens \\
REAL DB (Enamine)\citep{shivanyuk2007enamine} & 1.95 B &Synthesizeable compounds M$\leq$500, Slogd$\leq$5, HB$\leq$10, HB$\leq$5 \\ 
& &  rotatable bond$\leq$10, and TPS$\leq$140\\
ZINC15\citep{para_vs25}& 980 M&  Synthesizeable, available in ready-to-dock format \\
Pubchem\citep{kim2016pubchem} & 90 M & Literature derived bioactive compounds \\
Chemspider\citep{williams2010chemspider} & 63 M& Curated database with chemical structure and physicochemical properties \\
ChEMBL\citep{gaulton2012chembl} & 2 M & Manually curated drug-like bioactive molecules \\ \hline
\end{tabular}
\label{topchemical}
\end{table}

\section{Milestones in virtual screening} 
\label{miles_sec}
The first virtual screening using 3D structures of chemical compounds was carried out in 1990 against the target, dopamine D2 agonists based on which agonist with pKi 6.8 was successfully found~\cite{para_vs13}.  A similar search against other targets such as alpha-amylase, thermolysin, HIV-I protease yielded inhibitors with significant inhibition potential~\cite{para_vs13,para_vs14}. The top compounds obtained from screening were found to be potential inhibitors and with further optimization of the lead compounds, the inhibition potential increased considerably. With the use of more accurate scoring functions and large sized chemical libraries and powerful computers, the computer aided drug discovery can  become a potential route to narrow down the search space before subjecting to experimental high throughput screening. Thanks to the currently available Petaflop/s computing facilities, one can screen billion compounds in a day. The latest record on the high speed virtual screening is reported by Jens Claser et al., where the authors screened a billion compounds within 21 hours~\cite{para_vs15}.

The number of compounds screened against various targets keeps increasing with time. As the chance for a compound with better binding affinity increases with size of the chemical library such screening procedures result in identifying highly potent compounds. We here list a few example cases (refer to Table 2.): The screening study by J. Lyu et al. yielded an active compound (339204163), which had 20 times more potency than the known inhibitors for AmpC target~\cite{para_vs16}. The most potent inhibitor for the same target was designed by the same research group by optimizing the lead compound (the end compound was referred to as 549719643). Similarly 10 fold more potent agonists (465129598, 270269326, and 464771011) were identified for the D4 receptor~\cite{para_vs16} . The most potent compound with 180 pM binding affinity, 621433144 for the same target has been identified by the same group. Using the multistage docking workflow, referred to as Virtual Flow for Virtual Screening (VFVS)  (where different docking softwares  such as Quickvina2, Smina vinardo, Autodock Vina were used in sequence), recently the inhibitors for KEAP1 target were identified from the chemical space of 1.4 Billion compounds (made of Enamine Real database and ZINC15)~\cite{para_vs17}.  The first round of scoring was carried out using Quickvina2 while the rescoring was carried out for the top 3 Million compounds using Smina vinardo and Autodock Vina. An inhibitor, iKEAP1 with dissociation constant 114 nM was identified which is shown to interrupt the interaction between the KEAP1 and transcription factor NRF2~\cite{para_vs17}.
 
Thanks to parallel implementations of molecular docking softwares not only the number of compounds used for screening increased drastically but also the time required to accomplish the high throughput screening reduced considerably.  The inhibitors for the targets Purine Nucleoside Phosphorylase and Heat Shock Protein 90 were identified from the REAL enamine database (1.4 B compounds) using GPU enabled Orion software ~\cite{para_vs18}.  The recent screening of compounds from Enamine database against the enzyme targets from SARS-CoV-2 has been carried out in 21 hours instead of 43 days with the use of parallel implementation of Autodock4.0 (referred to as Autodock-GPU) on the Summit supercomputer~\cite{para_vs19,para_vs20}. 
Currently the largest experiment ever run has been done using the EXSCALATE platform based on LIGEN software. It virtual screened a library of $71.6$ B compounds against $15$ docking sites of $12$ viral proteins of SARS-CoV-2.
The experiment has been carried out on the CINECA-Marconi100 and ENI-HPC5 supercomputers, and it lasted $60$ hours~\cite{bigrun} performing overall more than 1 trillion docking evaluations.
A list of megadocking and gigadocking screening calculations on large sized chemical libraries is presented in Table \ref{table2}.

\noindent
\begin{table}[]
\scriptsize
    \centering
     \caption{Topmost mega or gigadocking applications reported in the literature}
\begin{tabular}{lllll} \hline
{\bf NO} & {\bf Year} & {\bf Target} & {\bf No of compounds} & {\bf Docking tool}\\ \hline
 1 & 2019&enzyme AmpC & 99 M & Dock3.7 \\ 
  2 &2019&D4 dopamine receptor & 138 M & Dock3.7\\ 
  3 &2019& Purine Nucleoside Phosphorylase  &1.43 M& Orion\\ 
  4& 2019& Heat Shock Protein 90 &1.43 M & Orion \\
  5&2020&KEAP1 & 1.4 B&  Quickvina2\\ 
 6 &2021& Mpro & 1.37 B& Autodock-GPU \\
 7 &2021& 12 SARS-CoV-2 Proteins & 71.6 B & LiGen \\ \hline
\end{tabular}
\label{table2}
\end{table}

\section{High-Performance Computing}
\label{hpc_sec}
Molecular docking for a single chemical compound involves sampling and scoring. This refers to sampling over the configurational phase space for the compound in the target binding site and computing the scoring functions for each of these configurations generated. In the case of virtual screening, compounds from a chemical library are ranked against a single target with respect to their binding affinities and this additional component is referred to as ranking (refer to Figure \ref{figure_cadd}).  As discussed in the previous section this is computationally very demanding and can be highly benefited from the use of the parallel algorithms which can run on  high performance computers with different (shared memory and/or distributed) architectures~\cite{para_vs11,para_vs12}.  Nowadays, the massively parallel computing units referred as graphical processing units (GPUs) and Field Programmable Gate Arrays (FPGAs) are accessible to research groups or even individuals and with the development of parallel virtual screening softwares drug discovery projects can be offloaded to such groups making the drug discovery economically sustainable. Below we  highlight the opportunities for the parallelization of VS protocol.

\subsection{Parallelization Strategies of Virtual Screening for High-Performance Computers}
As we discussed above the virtual screening involves three key steps: 
\begin{enumerate}[label=(\roman*)]
\item Sampling over configurational phase space of ligands within the binding site 
\item Estimating the scoring function for each of the configurations of the chemical compound within the target binding site to identify the most stable binding mode/pose.
\item Ranking of compounds with respect to their relative binding potentials
\end{enumerate}
Each of these key steps can be parallelized as the computations can be carried out independently. 

The estimate of scoring function for each of the binding mode/pose within the protein binding site involves the computation of docking energies or binding free energies or any other empirical potentials. We need to find the most stable binding mode for each of the ligands in the target binding site (which corresponds to a global minimum in the protein-ligand potential energy surface). So, for each ligand millions of configurations (each configuration is a point in the ligand configurational phase space) are generated and the scoring functions are estimated for these structures. As we discussed above, the configurations can be generated by changing the translational, rotational or torsional degrees of freedom. These changes can be performed with a deterministic methodology, or by using random-driven approaches such as Monte Carlo simulations or Genetic Algorithms. The estimate of energies for each of these configurations can be carried out independently, these can be distributed to different computing units.

\begin{figure}[t]
\centering
    \includegraphics[width=0.75\textwidth]{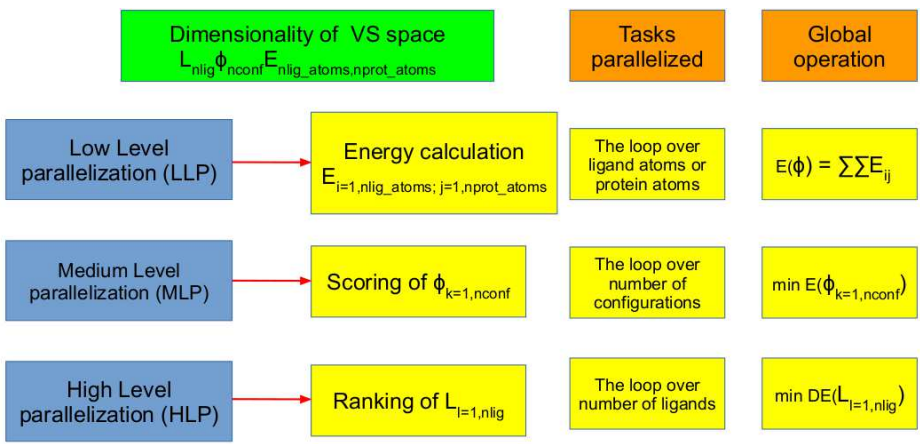}
    \caption{Various parallelization opportunities in virtual screening. The terms nlig, nconf, nlig\_atoms and nprot\_atoms refer to number of ligands in a chemical library, number of configurations (specific to each ligand and dictated by number of rotatable bonds), number of ligand atoms and number of protein atoms respectively.}
    \label{figure_parallelism}
    \end{figure}

In the virtual screening the aforementioned procedure is repeated for all the ligands in the chemical library and even this can be carried out independently and so can be distributed to different computing units. 
Finally, even the calculation of the energy of a single conformation can be parallelized since it requires to estimate the interactions of all the ligand-protein atoms couples. Overall, the parallelization of the virtual screening approach can be implemented in the following three steps, as we also show in Figure \ref{figure_parallelism}. 

\begin{enumerate}[label=(\roman*)]
\item \textbf{Low Level Parallelization (LLP)}: Parallelizing the energy calculation.
\item \textbf{Mid Level Parallelization (MLP)}: Parallelizing the conformer evaluations and scoring.  
\item \textbf{High Level Parallelization (HLP)}: Parallelizing the ligands evaluations on different computing units.  
\end{enumerate}

These three different levels have different efficiencies, since the amount of data transfer between the computing units is very different and may become a strong overhead.
Moreover, it is also possible to combine more of these strategies in a single application to address different level of parrallelism available (such as, for example, HLP for multi-node and MLP for multi-core architecures).
The low level parallelism approach is however usually not appealing, since the computed docking energies for different docking poses need to be compared and this involves frequent data transfer between different computing units. 
The docking energy calculation involves summation over the pairwise interactions of the atoms centered on protein and ligand subsystems. 
In case of flexible targets the intramolecular energies as well need to be computed which is obtained from a double summation over the number of atoms in the target. 
The non-bonded interaction calculation (sum of electrostatic and van der Waals) is the most time consuming part of the energy calculation (by 80\%)~\cite{para_vs11}.  Stone et al. showed 100 times speed up for the non-bonded energy calculation using GPUs while Harvey et al. showed 200 times increased performance~\cite{para_vs26,para_vs27} . 
A more sophisticated algorithm which effectively distributed the tasks to CPUs and GPUs  for computing non-bonded interactions was developed by Gine’s D. Guerrero et al.~\cite{para_vs28}.  In this case, each atom of the receptor was assigned to a single thread in GPUs which handles computing its interaction with all the ligand atoms~\cite{para_vs28} . Each thread was provided with necessary ligand and receptor atom coordinates and charges. The speed up due to this algorithm was up to 280 times.  
 
The parallelization of scoring function calculation is the most problematic approach, since it requires to sum all the atom contributions that are evaluated by different compute units (see Figure \ref{figure_parallelism}).
This introduces frequent small data transfers that may be limiting the scaling behavior
For this reason, the other two techniques are the most used in parallelizing VS software.
Indeed, computing the energy for different configurations of the ligand can be carried out independently. Similarly finding the global minimum structure for each of the ligands as well can be carried out independently. 
Each molecular docking step involves generation of millions of configurations and the scoring functions for all the conformers can be calculated independently by assigning the tasks to different computing units. 
The energies for conformers can be gathered and checked for the least energy configuration corresponding to global minimum in the protein-ligand free energy surface. 

\subsection{HPCs and Accelerators technology as problem solver}

Parallel computing architectures can be classified based on the memory availability to computing units:
\begin{enumerate}[label=(\roman*)]
\item In the shared memory architectures, a set of processors use the same memory segment 
\item In the case of distributed memory architectures, each computing unit has its own memory.  
\end{enumerate}

In general a high performance computing environment is usually made of both architectures where the shared memory multi core CPUs are connected through high-performance network connections. 
The memory in different nodes are local and are available to those cores within the specific node. The parallel execution of tasks in a shared memory architecture can be achieved by  compiler directives such as OpenMP pragmas or libraries such as pthreads.

On the other hand, the parallel implementation of tasks in distributed memory architectures requires to handle the communication manually, using libraries such as MPI (Message Passing Interface) between different nodes of the machine.

Since most of the supercomputers have both shared memory behavior (on a single node) and distributed memory behavior as they are multi node machines, it is easy to see the usage of both programming libraries for parallelizing the VS.
In addition to multiple CPUs, recent HPCs also are integrated with accelerators such as graphical processing units which have their own memory unit. 
The programming languages for CPU+GPU supercomputers are CUDA, OpenACC and OpenCL.
To use the accelerators, it is mandatory to move the data between the host and the device memory, which means that the developer needs to copy the data on the device before performing the computation and the result from the device after the computation is done.
In addition, the tasks distribution to GPU cores and synchronization of the task execution in CPU and GPUs are achieved.

In virtual screening, the docking of different ligands can be carried out independently from each other. The procedure can be effectively parallelized on supercomputers with thousands of multiprocessor nodes as well as in multi CPU+GPU computers. As we discussed above there are different possibilities for running the virtual screening in HPCs. The molecular docking procedure for each ligand can be parallelized by distributing the docking energy calculations for different conformers of the ligand in a target binding site. In this case, the receptor coordinates and ligand coordinates and their charges need to be provided to all computing units (i.e. server nodes or GPU cards). In the second scenario, the molecular docking workload for different ligands is distributed to different computing units. In this case, the receptor coordinates and charges should be made available to all the computing units while the ligand coordinates can be made available to the specific computing unit handling this specific ligand. The main disadvantage in this way of distributing tasks is that the ligands can have different sizes and so a rank assigned with the smaller sized ligand completes the task and waits until the molecular docking procedure is completed for all the ligands in different ranks to accept the tasks of second round screening. This way the computing time in this specific node is wasted. So, it is usually recommended to sort the ligands in terms of their size and then the distribution of ligands can be carried out from the list. In this way, all the ranks will have more or less equivalent sized tasks and the wall time in different ranks is efficiently used. The next section provides an overview of the different software targeting HPC systems also highlighting the type of parallelization employed.

\section{Current implementation of VS available for Workstations, accelerators and HPCs}
\label{pvs_sec}
Currently there are many parallel implementations for doing virtual screening in multicore machines, clusters and accelerators. As we discussed above, different strategies were adopted for parallel virtual screening. Only a few virtual screening softwares such as Flexscreen use the parallel implementation of docking energy calculation. The remaining do the parallelization by distributing the conformational sampling/scoring or the ligands ranking segments over different computing units. 

Table \ref{tab:comparison} shows a comparison of the analyzed software under some key features.
As we can notice, most of the available software have the possibility to scale on multiple nodes, thus are able to exploit the whole HPC machine.
All of them are able to fully utilize a node, while only a few have the support for the GPU acceleration.
Moreover, as we already anticipated, all of them use a MLP (parallel conformational search) or HLP (parallel ligand evaluations) or both strategies to parallelize the computation, while none of them uses a low level parallelism (parallel energy evaluations).

We will now analyze one by one more in detail these softwares: we will focus on their performance in massively parallel architectures and accelerators when compared to CPU implementation. Wherever possible the accuracy of the docking results (in terms of reproducing the experimental protein-ligand complex structure and experimental inhibition constants) will be discussed.
\begin{table}
    \caption{Different parallel virtual screening softwares and important features} 
   \label{tab:comparison}
   \tiny
\begin{tabular}{p{1.5cm}p{3cm}p{2.5cm}p{2.0cm}p{2.0cm}p{0.5cm}p{0.5cm}p{0.5cm}}
{\bf Software} & Parallelization  & Programming & Scoring & Minimization &Multi &Multi  & GPU \\
& segment  & Language & function &  & Thread& Node& \\ 
\vspace{0.4cm}\\ \hline
\vspace{0.2cm}\\
Dock 5,6 & Conformational search & C++, C, Fortran77, MPI & Physics based and Hybrid & Monte Carlo &Yes & Yes & No \\ \\
Autodock Vina & Conformational search & C++, OpenMP & Hybrid & Monte Carlo &Yes & No & No \\
MPAD4 & Ligand screening, Conformational search & C++,  MPI, OpenMP & Physics based & Lamarckian GA & Yes & Yes& No \\

VinaLC &  Ligand screening, Conformational search
&  C++, MPI, OpenMP& Hybrid & Monte Carlo & Yes & Yes & No  \\

VinaMPI &  Ligand screening, Conformational search
&  C++, MPI, OpenMP& Hybrid & Monte Carlo & Yes & Yes & No \\ 

Ligen Docker-HT & Ligand screening, Conformational search & C++, MPI, CUDA & Empirical & Deterministic & Yes & Yes & Yes \\

GeauxDock & Ligand screening, Conformational search & C++, OpenMP, CUDA & Physics and Knowledge based &  Monte Carlo& Yes & Yes & Yes \\

POAP &  Ligand screening & bash &  Same as parent Docking software& Same as parent Docking software&  Yes & Yes &  No \\

GNINA &  Conformational search & C++ & Empirical and CNN ML &  Monte Carlo & Yes& Yes &Yes \\ 

Autodock-GPU &  Ligand screening & C++ and OpenCL & Physics based &MC/ LGA &  Yes & No & Yes \\  \hline
\end{tabular}

   \end{table}

\subsection{Dock5,6}
Dock5 and 6\citep{dock5,dock6} are the parallel implementations of virtual screening program written in C++ with MPI libraries. They are based on the original version referred to as Dock1 which was developed by Kuntz and coworkers\citep{dock1}. The Dock1 used a geometric shape matching approach for identifying the lead compounds for a given protein target. The subsequent versions adopted physics based scoring functions for ranking and had improvement over the thoroughness of sampling and accounted for the ligand flexibility. The recent version allowed using multiple scoring functions where the solvation energies are computed using different implicit models such as Zou GB/SA,\citep{zou_gbsa} Hawkins GB/SA\citep{hawkins_gbsa}, PB/SA\citep{pbsa} and Generalized solvent models as implemented in Amber16. The benchmarking study to evaluate the performance of recent Dock6 (V6.7) used SB2012 dataset\citep{sb2010_db} which is a collection of crystallographic structures for about 1043 receptor-ligand complexes. It was able to reproduce the crystallographic poses in this dataset to an extend of 73.3\%  (i.e. RMSD between the experimental and predicted binding poses was $<$2 \AA~) which is due to reduction in the sampling failure of previous Dock versions. There is a also report in the literature on the offloading of Dock6  to CPU+GPU architectures using CUDA.~\citep{dock6_gpu} Only the ranking using amber scoring was offloaded to GPU architectures. In this offloading, the coordinates, gradients and velocities are copied to host (CPU) memory to device (GPU) memory and the results are copied back from GPU to CPU memory. Since the GPU could only handle single precision numbers, the original data in double precision were converted to single precision numbers before transfer.  Overall, the study reported about 6.5 speed up for the amber scoring in Dock6 in GPU (NVIDIA GeForce 9800 GT) when compared to AMD dual-core CPU.~\citep{dock6_gpu}

\subsection{Autodock Vina}
Autodock4.0 is the most widely used molecular docking software based on physics based scrored function but the original version is sequential in nature.  However the Autodock Vina, which is also from the same developers at Scripps institute, can use multiple cores simultaneously for carrying out the docking. The calculations can be executed with single threaded or multithreading options. The implementation is using C++ with Boost thread libraries. In the multithreading version, mutiple Monte Carlo simulations are initiated with different random number seed to explore different areas of conformational space of the ligand within the binding site.  In a benchmarking study, for the same protein-ligand complex, the Autodock Vina~\cite{para_vs9} with single thread ran 62 times faster than the Autodock4.0. Further running Autodock Vina with multi threading option in 8 CPU machines yielded 7.3 times faster (when compared to single threaded option) completion of molecular docking. The performance of single threaded Autodock Vina compared to Autodock should be attributed to the difference in the computational cost of scoring functions. The more than 7 times speed up with multi-threading option in an eight core machine shows the molecular docking calculation scales well with the number of cores. Autodock vina relies on OpenMP for distributing tasks to different threads and so is suitable for workstations with multiple cores. However for the supercomputers with distributed memory, this version of Autodock Vina is not suitable and rather more robust programs that allow the data transfer and communication between different nodes need to be used.  It is also worth mentioning that there are updated versions of Autodock vina namely Qvina 1 and Qvina 2, which showed some speedup due to the improvement in local search algorithm. SMINA~\citep{smina} is a fork of Autodock vina with a number of additional features such as user specified scoring function,  creating grid box for docking based on the coordinates of ligand bound to target, improved minimization, feasibility to include residues for flexible docking, possible to print more than 20 poses. In terms of speed up in HPCs, this did not contribute to any improvement. 

\subsection{MPAD4}
MPAD4~\cite{para_vs30}  is a parallel implementation of Autodock4.0 and the important features when compared to parent code are listed below: i) It uses MPI to distribute docking jobs across the cluster ii) The grid maps generated for receptor are reused for all the docking calculations while in the default version these files are generated for each ligand docking with the target receptor and loaded into memory and released at the completion of docking. In MPAD4, The maps are loaded into memory of the node at the beginning of tasks and are used for all remaining docking calculations. This greatly reduces I/O usage contributing to speed up in the performance.  iii) The OpenMP is used for the node level parallelization in executing the LGA for finding the global minimum.  Overall it allows system level and node level parallelization. The performance analysis of MPI version and MPI+OpenMP version can be studied by setting OMP to 1 and 4 respectively. The computers used for the initial performance analysis were IBM Blue gene/P and shared memory 32 core POWER7 p755 server. The dataset used for performance analysis was HIV protease target and 9000 compounds from a druglike subset of ZINC8 database (which has 34481 ligands in total). The performance analysis in Blue gene/P with 2,048 (8,192) node (core) showed that Grid map reuse has reduced single threaded execution time by 17.5\%. Multithreaded execution of the code yielded 25\% improvement in the overall performance.  The execution of the code on nodes ranging from 512- 4096 showed near linear scaling behavior for symmetric multi processing (SMP) mode (OMP=4). In particular, for 16384 core system, the speed gained was 92\% to that of the ideal case.  The virtual node (VN) mode (OMP=1) with the grid map “reuse” option showed however 72\% speed up when compared to the ideal case. The gain in the performance of SNP mode should be attributed to multithreading. The node utilization can be further improved with the use of pre-ordering ligands with decreasing number of torsional angles.   

\subsection{VinaLC}
VinaLC~\cite{zhang2013message} is an Autodock Vina extension to use MPI library for parallelization in large supercomputers. This implementation is very similar to VinaMPI (which is described below) and uses MPI and multithreading hybrid scheme for parallelization across nodes and within node respectively. The computational cost of each docking calculations depends on the size of the ligand, receptors and the grid box. If the calculations are distributed on all MPI processes due to this uneven size of the input systems, there can be MPI load imbalance where the processes are waiting for other processes to complete the task. This load imbalance is tackled effectively by the Master-Slave MPI scheme where the master process handles the inputs, outputs and job allocation to the slave processes. The tasks in the master slaves are handled by 3 for loops: (i) the first loop is over each combination of receptor target-ligand which assigns docking task to a free slave. (ii) The second loop collects the docking results from slave processes and the new tasks are assigned in case of unfinished calculations (iii)  The third loop frees the slave processes. In the slave processes an infinite while loop is initiated which ends when the "job finished" signal is received from the master.  The ideal slave processes are identified from the MPI\_ANY\_source tag and docking tasks are assigned upon the completion of previous task. In this way, the computing resources available in different processes are utilized efficiently. To make the communication effective all the inputs needed for a single docking calculation are sent by single MPI\_Send call. So, coordinates of the receptor and ligand and grids (which are computed on the fly in the master process) are sent to the slave process. The outputs from the slave processes are collected by the master process into a few files in stead of generating file for each ligands (which will generate million or billion files depending upon the size of the chemical library). The benchmarking study was carried out using the two datasets namely ZINC and DUD (directory of useful dacoys). The target was chosen as Thermus
thermophilus gyrase B ATP-binding domain. The benchmarking calculations were carried out on HPC machines at Lawrence Livermore National Laboratory and the number of cores used were in the range 600-15408. The study showed that the average CPU time per docking was closer to ideal average CPU time. The VinaLC was shown to scale well up to 15 K cores. The percentage of I/O activity was reported to be negligible when compared to the total computing time. For aforementioned target using 15 K cores, VinaLC could screen about one million compounds from Zinc15 database in 1.4 hours. This can be extrapolated to 17 million compounds per day which suggests the suitability of VinaLC for the most time taking mega or gigadocking screening applications.      

\subsection{VINAMPI}
VinaMPI is another implementation of Autodock Vina for distributed computing architectures~\cite{para_vs29}. It is written in C and for communication between the nodes, it uses MPI libraries. In order to avoid poor scaling behavior of the parent-child (or master-slave) distribution scheme in massively parallel supercomputers, this implementation uses all-worker scheme. It is worth recalling that rather VinaLC used Master-slave scheme for distributing tasks.  In this code, each worker (or each MPI rank) deals with its own protein-ligand complex and within each rank the computation (related to search of global minimum) is carried out using multithreading. Due to this reason it is also suitable for the virtual screening for more than one targets. Further the computations are sorted out in terms of complexity so that the work loads at a given round of screening can be distributed equivalently. The computational complexity is measured based on the number of rotatable bonds and size of the ligands which is used to sort the tasks in the beginning of the screening. As a benchmark, the dockings were carried out for targets namely ACE (angiotensin-converting enzyme), ER AGONIST (estrogen receptor agonist), VEGFR2 (vascular endothelial growth factor receptor kinase), and PARP (poly(ADP-ribose) polymerase) with a chemical library of  98,164 compounds (comprised of ligands and decoys). Running this screening in a 516 cores supercomputer costed about 103 s~\cite{para_vs29}. It is expected that the same implementation in a supercomputer with 0.3 M cores can be used to screen 250 M compounds in 24 hours.

\subsection{LiGen Docker-HT}
LiGen~\cite{beato2013use} is a VS software that leverages CPU and GPU to perform the required computation.
Several versions of the tool have been developed, starting from a CPU only application~\cite{beccari2013ligen,beato2013use,tunableapprox}, then the main kernels have been ported to the GPU by ~\cite{vitali2018accelerating} using OpenACC. Finally, it has been optimized using CUDA for the GPU kernels and this last version has been used to perform a large VS experiment in the search of a therapeutic cure for COVID-19 screening 71.6B compounds against 15 binding sites from 12 Sars-Cov2 viral proteins on 2 supercomputers accounting for 81PFlops~\cite{bigrun}.
LiGen uses deterministic algorithms to generate the different conformers of a ligand, and an empirical scoring function to select the best molecule.
The Docker-HT application is the version of LiGen that is designed targeting a large VS campaign, and it is able to leverage multi-node, multi-core and heterogeneous systems.
In particular, it uses MPI to perform the multi-node communication, which is limited as much as possible by the algorithm to avoid large communication overheads~\cite{pdswmarkidis}.
Indeed, the amount of data that needs to be processed by every node is divided beforehand, and it may create load balancing issues since there is no mechanism to re-balance it during the execution of the application.
On the single node, it leverages the C++ thread library.
Finally, it uses CUDA to support GPU acceleration. Within each node, the program uses pipeline parallelism and work-stealing to process the ligands.
   
\subsection{Geauxdock}
This is a parallel implementation of virtual screening available for multicore CPU, GPU and Xeon Phi-computers. The software uses a common code for front-end computations in all these computers~\cite{para_vs36}. However, the back end codes have one version for CPU, Xeon Phi architectures and another for GPU. The code for CPU and Xeon Phi architectures written in C++, OpenMP and IntelSIMD pragmas. The GPU version is written in C++ and CUDA. The program uses the Monte Carlo approach for conformational search and for identifying the global minimum of the protein-ligand complex. The scoring function is based on physics based energy terms combined with statistical and knowledge based potentials. The performance of the code has been tested in multicore CPUs and massively parallel architectures namely Xeon Phi and NVIDIA GPUs. The testing using CCDC/Astex dataset showed 1.9 times increase in performance for Xeon Phi when compared to 10-core Xeon CPU. Further on the GeForce GTX 980 GPU accelerator, the performance was 3.5 times higher when compared to the CPU version.   

\subsection{POAP}
Poap is a GNU parallel based multithreaded pipeline for preprocessing ligand, for doing virtual screening and for post processing the docking results~\cite{para_vs31}. It also allows the minimal use of memory through optimized dynamic file handling protocol. It has also been optimized in a way that erroneous ligand input does not affect the workflow. It can be integrated with any of other sequential or multithreaded molecular docking softwares such as Autodock4.0, Autodock Vina or AutodockZn. In the case of Autodock based virtual screening, the map files are generated for each ligand in the datasets. In the case of POAP, the map files are directed to a common hub directory and so occupancy of space due to redundant atom types in ligands is overcome. The number of threads to be used should be mentioned when the Autodock vina is used as a docking software. In the case of Autodock and AutodockZn, the number of parallel jobs to be executed (which can be equal to the number of CPU threads) should be defined by the user. The performance of POAP has been tested using the virtual screening for the targets namely Human ROCK I, HTH-type transcriptional regulator, Polyketide synthase and PqsA (Anthranilate-coenzyme A ligase) using the ligands from chemical library, DrugBank.  Since POAP does not have any serial code, the speed up (theoretical estimate) is directly proportional to the number of processors used. The performance analysis of Autodock in a T5510 DELL workstation with Intel Xeon(R) CPU E5-2620V2, 2.10 GHz clock speed (12 Cores, 24 threads) with 62.9GB RAM showed 12.4 times speed up when compared to serial mode (the number of parallel jobs specified is 24). Similarly, the Autodock Vina showed 2.4  times speed when compared to default mode (which is already multithreaded) and here the number of jobs was set to three.

\subsection{GNINA}
Gnina is a fork of SMINA~\cite{smina} and Autodock Vina~\cite{para_vs10}. When compared to the hybrid scoring function employed in Autdock Vina, it provides options to use various built-in scoring functions (such as Vina, Vinardo) along with user customized scoring  functions. More importantly, it provides a machine learning based scoring function to rank the complexes. The default scoring function (called by “none” option) is the same as used in Vina or Smina while the rescoring (called by “rescoring” option) allows the top most ligand poses to be ranked using machine learning based scoring functions. 

In particular, this scoring function is based on convolutional neural networks trained using 3D protein-ligand complex structures (as reported in PDBBind or BindingDB) and corresponding experimental inhibition constants. They were trained to reproduce binding pose and the binding affinity. There are multiple machine learning functions (namely crossdock\_ default2018, dense, general\_default2018, redock\_default2018, and default2017) provided by GNINA and these have been developed using different datasets. The CNN based scoring function outperforms the scoring function implemented in Vina in reproducing the binding poses. The RMSDs for the predicted poses  in the unseen examples are below 2 Å in as many cases as 56\%. Further the binding prediction of poses within this cutoff improved to 79\% if the redocking is performed.  

When compared to Autodock Vina, the grid box center can be provided with the help of a ligand file. For the conformational search, GNINA uses Monte Carlo sampling scheme. The sampling is carried out over the ligand translational, rotational and torsional degrees of freedom. In the case of flexible docking, the sampling is also carried out over the residue side chain conformations.  Unlike Smina, Gnina does computing in single precision (32 bit) which allows the possibility of offloading the CNN scoring tasks  to GPUs. Even though Gnina can be used in massively parallel supercomputers and HPCs with accelerators, there is no performance analysis or profiling when compared to other docking softwares reported in the literature.

\subsection{AUTODOCK-GPU}
Autodock4.0 is one of the most widely used molecular docking software but it is a serial code which runs on a single thread so can not be effectively used in high performance computing environments with multiple CPUs and GPUs. Autodock-GPU~\cite{para_vs33} is the version of autodock developed for multiple node parallel computers with GPU accelerators. It is worth recalling that the above discussed MPAD4 was developed for multi CPU architectures. This program has been developed using the application programming interface, OpenCL as it allows portability to hybrid platforms with CPUs and GPUs. When compared to Autodock4.0, the local search algorithm uses derivatives of energies with respect to translations, rotations and torsions (this implementation of gradient based local search is referred to as ADADELTA). In the case of CPU+GPU architectures, the workflow consists of a sequence of host and device functions. In analogy to biological gene, the state of the protein-ligand complex is represented by a sequence of variables. In the case of a rigid docking (where the protein framework is treated as a rigidbody), the variables represent positions, orientations and conformation of the ligand. The number of variables are 6+Nrot where Nrot is the number of rotatable bonds. The docking is aimed at finding the genotype which corresponds to the global minimum in the protein-ligand potential energy surface and the ranking is dictated by the scoring function. 

The performance of Autodock-GPU has been tested using the diverse data set of 140 protein-ligand complexes from Astex Diversity Set~\cite{para_vs34}, CASF-2013~\cite{para_vs35} and protein databank. The reference docking calculations were performed using the single threaded Autodock4.2.6. The speedup performance was dependent on the minimization algorithm used for the local search (whether Solis-Wets or ADADELTA), GPU type and the type of protein used in the docking.  With the use of Solis-Wets local search algorithm, the speed up was 30 to 350 times in GPUs with the M2000 showing the least performance and with TITAN V showing the best. However, with the use of  ADADELTA local search, the speed up was only 2 to 80 times improved which has to be attributed to the computationally expensive calculation of gradients and difficulties associated with the parallelization of this local minimization step. In general, TITAN V cards showed 10 times higher speed up when compared to M2000 versions. The performance analysis in multiple core CPUs showed a similar trend where for the  Solis-Wets search the speed up was in the range 5 to 33 times (the number of cores employed 8-36) while for the ADADELTA local search the speed up was 2-20 times better. 

\begin{table}[]
\caption{Timeline for different parallel virtual screening software and source URLs}
    \centering
   \footnotesize
\begin{tabular}{llll}
\vspace{0.05cm}\\
No&Year & Parallel VS & source\\ \hline
\vspace{0.05cm}\\
1 & 2006 & Dock5\&6 &http://dock.docking.org/ \\
2&  2009 & Autodock Vina   &  http://vina.scripps.edu/\\
3 & 2011 & MPAD4 & http//autodock.scripps.edu/downloads/multilevel-parallel-autodock4.2 \\
4&  2013 & vinaMPI   &  https://github.com/mokarrom/mpi\-vina\\
5 & 2013 & vinaLC & https://github.com/XiaohuaZhangLLNL/VinaLC\\
6 & 2016 &GeauxDock & http://www.brylinski.org/geauxdock\\
7 & 2018 & POAP & https://github.com/inpacdb/POAP\\
8 & 2021 & Autodock-GPU &  https://github.com/ccsb-scripps/AutoDock-GPU \\
9 & 2021& GNINA& https://github.com/gnina/gnina  \\ \hline
\end{tabular}
\label{table_timeline}
\end{table}

\subsection{Other VS tools}
The focus of this review was mostly about the open source parallel VS softwares which are summarized in Table 3 along with some important features. The details about the year they were introduced and source URL pages are listed in Table \ref{table_timeline}.  Many of these softwares such as GeauxDock, Autodock-GPU, GNINA, are introduced in the recent years and so there capacity in the lead compounds identification from huge chemical libraries needs to be validated extensively. Meanwhile, many already existing virtual screening softwares have contributed to successful lead compound identification and lead optimization over the years. In particular, the softwares such as FlexX, DOCK (the sequential version of above discussed Dock6), SLIDE, Fred (OpenEye), GOLD, LigandFit, PRO\_LEADS, ICM, GLIDE, LUDI and QXP are worth mentioning~\citep{lyne2002structure}. Among these, LigandFit, QXP employ Monte Carlo for sampling while SLIDE and Fred employ conformational ensembles approach. GOLD, ICM and GLIDE respectively adopt Genetic algorithm, pseudo-Brownian sampling/local minimization and exhaustive search for sampling. Finally Dock and FlexX use incremental build approach for identifying the most stable binding mode/pose for the ligand. There are other VS softwares such as RosettaDock~\citep{rosettadock} and Surflex~\citep{surflex} which are not discussed here as the review focuses on those with parallellism capability.

\begin{figure}
	\begin{center}
		\includegraphics[width=0.75\textwidth]{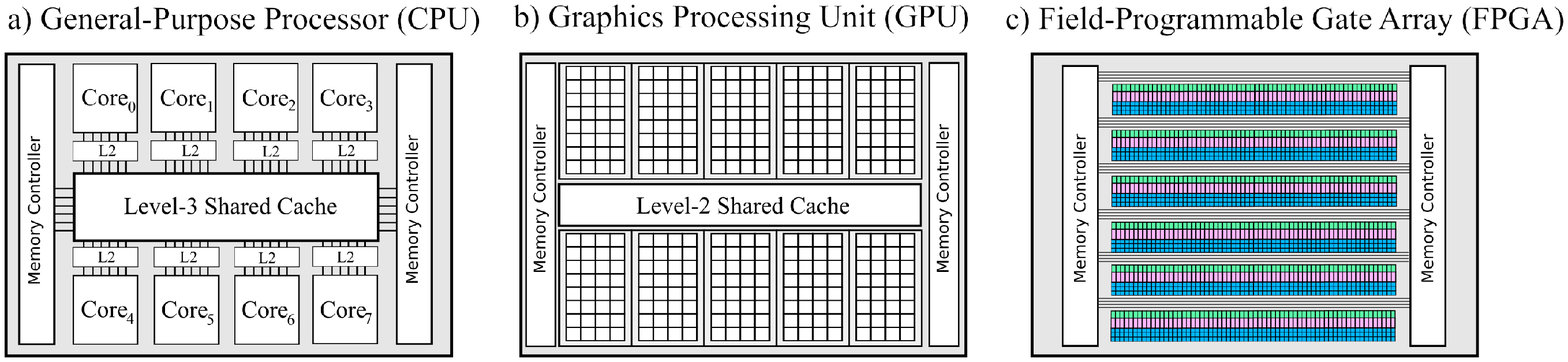}
     \caption{A conceptual picture over different processors and accelerators, showing (a) CPUs, which is a latency-focused architecture with few processing units and a large (and deep) memory hierarchies, (b) a GPU, which is a throughput-focused architecture with more processing units (contra CPUs) and a shallower memory hierarchy, and (c) FPGAs, which offer much more parallelism compared to both CPUs and GPUs and with finer control over individual unit types (here shown in different controls), but is harder to use.}
     \label{fig:figure_compute}
	\end{center}
\end{figure}

\section{Emerging Reconfigurable Architectures for Molecular Docking}
\label{fpga_sec}
In prior sections we have focused exclusively on reviewing methods of virtual screening and molecular docking that targets modern central processing unit (CPU) and graphics processing units (GPU) solution (refer to Figure~\ref{fig:figure_compute}). At the same time, we know that Moore’s law (transistor scaling) is terminating, which could motivate (or even necessitate) the search for alternative computing platform that can continue the performance trend that molecular docking has come to rely upon. Among the many (so-called) post-Moore technologies~\cite{vetter2017architectures}, reconfigurable architectures are perhaps the most noticeable, partially because they are readily available today. A reconfigurable architecture, such as an Field-Programmable Gate Array (FPGA) or Coarse-Grained Reconfigurable Array (CGRAs)~\cite{podobas2020survey} is a system which aspires to retain some of the silicon plasticity that is lost when manufacturing an Application-Specific Integrated Circuit (ASIC). 
In turn, users can leverage reconfigurable systems to perfectly match the hardware to the application, which in turn can lead to improvement in performance and reduction in energy costs. For example, the expensive von Neumann-bottleneck associated with the decoding of instructions in CPUs can be virtually eliminated. Traditionally, reconfigurable architectures such as FPGAs have been programmed using complex low-level hardware description languages (HDLs) such as VHDL or Verilog. This, in turn, has limited exposure of using these devices to hardware specialized and thus out of reach for the typical HPC users. However, with the increased in maturity of High-Level Synthesis (HLS)~\cite{nane2015survey} tools in the past decade, today programmers can describe their hardware in abstract languages such as C/C++ and directive-driven models (e.g., OpenCL~\cite{czajkowski2012opencl} or OpenMP~\cite{podobas2014accelerating}) and automatically translate the code down to specialized hardware. Modern HLS has, in turn, facilitated the accelerated use and research of FPGAs in other HPC applications such as (e.g.) computational fluid dynamics, neuroscience, and molecular docking.

Pechan et al.~\cite{pechan2010fpga} evaluated and compared the use of FPGAs against both GPU and CPU solutions of the popular Autodock software. They created a custom RTL-based 3-stage custom FPGA accelerator that computes the performance critical sections of the Autodock algorithm. More specifically, the custom accelerator has four modules capable of exploiting MLP (see section 4.1), while LLP is exploited inside each module (through pipelining); the accelerator relies on other methods to exploit HLP. They compared their solution against a custom (CUDA-based) GPU solution (GT220 and GTX260) and a CPU (Intel Xeon 3.2 GHz) version on the \texttt{1hvr} and \texttt{2cpp} protein pairs (from the Protein Data Bank). The overall results showed that FPGAs outperformed the CPUs for both use-cases independent on the number of dockings that were used. The GPU, however, had a clear advantage when a large number of dockings where executed, and the FPGA was only preferable when a few number of docking runs were executed. 
Recent work by  by Solis-Vasquez et al.~\cite{solis2018case,solis2019accelerating} focused particularly on using OpenCL HLS tools to create custom accelerators that targets FPGAs. Aside from disseminating their design-process, they also vary several different architectural properties in their accelerator. For example, they consider both floating-point and fixed-point representation for various phases of the computation, which demonstrates an advantage that FPGAs can provide over more general-purpose approaches. The accelerator runs at a fairly high frequency (between 172 MHz and 215 MHz) on a Intel Arria 10, and 
 consumes a  varying amount of resources (subject to their design-space exploration). They compare their accelerator against the single-threaded Autodock software on five protein targets, and show that they reach between 1.73x and 2.77x speed up.

Today, there is a remarkably small number of published work that leverage FPGAs in the Autodock software (for surveys using FPGAs on other molecular algorithms,  see~\cite{majumder2013hardware,pechan2012hardware}). What is even more surprising is that (to the authors' knowledge), CGRAs have been largely unexplored in this domain. With both FPGAs and CGRAs emerging as performance (and, more importantly, \textit{greener}) alternatives to traditional CPUs and GPUs, we believe that these systems will come to play a much larger role in molecular docking and virtual screening in the future than they have been so far.

\section{Conclusions}

The parallel implementations virtual screening algorithms in massively parallel computers with multiple CPUs and/or GPUs have the high potential to speedup the exploration of gigantic chemical spaces (having compounds in the range 10$^9$ to 10$^{12}$) in real time. In a serial version of the virtual screening softwares, it may take many years of CPU hours for such tasks. Current regard for gigantic docking is the screening of billion compounds from ZINC15 and Enamine database with the use of Autodock-GPU in Summit HPC computer in less than a day. The parallel implementations and reliable scoring functions will increase the success rates in the lead compounds identification for drug discovery. This makes the drug discovery less time consuming and economically sustainable. Further, as the chemical spaces are really huge the drugs with entirely different scaffold geometry can be identified.  The speed up of the virtual screening softwares is found to be dependent on the number of factors:  energy minimization algorithm, scoring function, biomolecular target and computer architecture. More elaborate studies will allow us to come up with highly optimized virtual screening softwares in the future. The implementation of VS for FPGAs is still in its infancy and a dedicated research is needed for adopting such architectures for drug discovery projects.





\bibliography{pvs_final}

\end{document}